\documentclass[11pt,preprint]{aastex}
\slugcomment{Accepted to {\it The Astronomical Journal}}

\def\kms{\ifmmode{\rm km\thinspace s^{-1}}\else km\thinspace s$^{-1}$\fi}
\def\ms{\ifmmode{\rm m\thinspace s^{-1}}\else m\thinspace s$^{-1}$\fi}

\newcommand{\teff}{\ensuremath{T_{\rm eff}}}

\newcommand{\re}{\ensuremath{R_{\rm e}}}
\newcommand{\me}{\ensuremath{M_{\rm e}}}

\shortauthors{Howell, et al.}
\shorttitle{$Kepler$ Small Exoplanet Host Stars}

\begin{document}

\title{Variability of $Kepler$ Solar-like Stars Harboring Small Exoplanets}

\author{
Steve~B.~Howell\altaffilmark{1} \\
NASA Ames Research Center, Moffett Field, CA 94035 \\
David R. Ciardi\\
NASA Exoplanet Science Institute, California Institute of Technology, 770 S. Wilson Ave., Pasadena, CA, 91125 \\
Mark~S.~Giampapa\\
National Solar Observatory\altaffilmark{2}, 950 N. Cherry Ave., Tucson, AZ, 85719 \\
Mark E. Everett\altaffilmark{1} \\
National Optical Astronomy Observatory, 950 N. Cherry Ave., Tucson, AZ, 85719 \\ 
David R. Silva\altaffilmark{1} \\
National Optical Astronomy Observatory, 950 N. Cherry Ave., Tucson, AZ, 85719 \\
Paula Szkody\altaffilmark{1} \\
Dept. of Astronomy, University of Washington, Seattle, WA, 98195
}

\altaffiltext{1}{Visiting Astronomer, Kitt Peak National Observatory, 
National Optical Astronomy Observatory, which is operated by the Association 
of Universities for Research in Astronomy (AURA) under cooperative agreement 
with the National Science Foundation.} 
\altaffiltext{2}{The National Solar Observatory is operated by the Association 
of Universities for Research in Astronomy (AURA) under cooperative agreement 
with the National Science Foundation.} 


\begin{abstract} 
We examine $Kepler$ 
light curve variability on habitable zone transit timescales for a large uniform
sample of spectroscopically studied $Kepler$ exoplanet host stars.
The stars, taken from Everett et al. (2013)
are solar-like in their properties and each harbors at least one exoplanet
(or candidate) of radius $\le$2.5\re. 
The variability timescale examined is typical
for habitable zone planets orbiting solar-like stars
and we note that the discovery of the smallest exoplanets ($\le$1.2\re) 
with corresponding transit depths of less than $\sim$0.18 mmag,
occur for the brightest, photometrically quietest stars.
Thus, these detections are quite rare in $Kepler$ observations.
Some brighter and more evolved stars (subgiants), the latter which often show large radial velocity jitter, 
are found to be among the photometrically quietest solar-like stars in our sample and the most likely small planet transit hunting grounds. 
The Sun is discussed as a solar-like star proxy to provide insights into the nature and cause of
photometric variability. It is shown that $Kepler's$ broad, visible light observations are insensitive
to variability caused by chromospheric activity that may be present in the observed stars. 
\end{abstract}

\section{Introduction}
The NASA $Kepler$ mission (Borucki et al., 2010) 
was launched in 2009 and completed four years of observation
of over 150,000 stars in a single field of view. $Kepler$ ceased science collection in
2013 after the failure of a second reaction wheel, disabling the ability to continue to 
precisely point at the original field of view. A new pointing strategy was developed and the
$Kepler$ mission was repurposed as the ecliptic viewing $K2$ mission 
starting in 2014 (Howell et al., 2014).
Currently, just over 1000 confirmed and over 4000 candidate exoplanets have been 
produced by the $Kepler$ mission.

The sample of stars observed by $Kepler$ consisted mainly of F to K stars 
with few M stars included in the initial target selection. Additional M stars were
added to the exoplanet star sample during the last two years of operation. 
The stars chosen were believed to mostly be normal, main sequence dwarfs generally covering 
a magnitude range of $V$=12-16, although the selection was not immune to containing
some evolved stars (subgiants) and indeed even
some giants to fill up the available observation list (Batalha et al., 2010; Brown et al., 2011). 
A number of $Kepler$ exoplanet host star follow-up studies, in particular
spectroscopic work (e.g., Huber et al. 2014; Marcy et al. 2014; Everett et al., 2013), have shown that some of 
the host stars are in fact slightly evolved, being closer to subgiants than luminosity class V stars.
This fact is particularly true for the brighter stars (V$<$12.5) more massive than the sun where 25-50\% are likely to be subgiants based on seismic modeling (Chaplin et al., 2014).
Over 50\% of the small ($<$2.5\re) exoplanets orbit stars fainter than R$\sim$14.5.
Since much of $Kepler's$ statistical power will come from small, potentially rocky planets orbiting 
faint stars, an understanding of their variability properties on transit timescales is warranted.

\section{The Solar-like {\it Kepler} Exoplanet Host Star Sample}

Everett et al. (2013) presented a uniform spectroscopic study of a large sample 
($>$200 stars) 
of faint $Kepler$ exoplanet host stars that harbor small ($<$2.5\re) exoplanets. 
These authors determined each star's effective temperature, gravity, metal content, and radius via 
rms minimization with model spectra and isochrone fits. 
The sample stars are very solar-like in temperature, 
spanning a relatively small range in effective temperature (approximately 2000K) 
centered on the Sun.

The Everett et al. (2013) stars represent a sample 
that should have much in common with our Sun. 
The sample consists of 220 stars characterized by effective temperatures from 4750 K to 7200 K (K4 to F3).  Most of the stars are in the apparent {\it Kepler} magnitude{\footnote{K$_P$ means "$Kepler$ magnitude" which is approximately $R$ band and defined in
Koch et al., (2010). Some stars in Everett et al. (2013) had unreliable or no "V" band data.
Appendix A provides a method to derive a V magnitude for any $Kepler$ star.} range of $K_p~$= 14.5 to 16 within the broad range for the entire $Kepler$ sample of $K_p~$= 12 to 16.
$Kepler's$ sample of candidate exoplanet host stars, culled from the
entire 150,000 stars observed in the single field of view, contains the majority of stars
at the faint end of the magnitude range,
a consequence of it being a magnitude-limited (magnitude selected) sample.
Thus, to statistically characterize the properties of a uniform sample of small planet host 
stars, it is important to understand the majority which are fainter than 14.5. 

Figure 1 presents the Everett et al. (2013) sample of 220 stars in the \teff-log g plane (a pseudo H-R Diagram) 
and shows the location of the zero-age and terminal-age main sequence (Schaerer et al. 1993) as well as the
location of the Sun. The ZAMS and TAMS lines were computed for solar metallicity and the details of the codes and opacities used are given in Schaller, et al., (1992).
Uncertainties in the determined \teff~ and log g values 
have been assessed in detail in Everett et al. (2013), yielding 
adopted (1$\sigma$ uncertainties) of 75 K for \teff~ and 0.15 for log g. 
These uncertainties, based on a sample of well studied bright stars, are on par with those reported 
by Torres et al., (2012) using similar spectroscopic fitting techniques but for brighter exoplanet host stars. 
The \teff~of this sample as a whole reveal it to be very solar-like, 
the range being only $\sim$1000K hotter or cooler than our Sun. 
Plotting the location of the relative density of occurrence for the ensemble of stars as they increase in 
temperature ($\sim$mass) in this figure illustrates the general evolutionary spread 
away from the Zero-Age Main Sequence into the subgiant region for the $Kepler$ exoplanet host stars.
It is likely that the percentage of subgiants to dwarfs in the $Kepler$ sample becomes
larger as the stars get brighter. For example, 
the Bastien et al. (2013) result that nearly 50\% of the $Kepler$ exoplanet host stars brighter than K$_P$=13 
are subgiants agrees with this expectation. 

\section{Variability Characterization}
\subsection{Kepler Quarter 9 Light Curve Sample}

Variability of exoplanet host stars can influence the transit survey completeness as increased photometric ``noise'' can hinder transit searches and detection algorithms.  In order to assess the variability of our sample of faint, small candidate exoplanet host stars, we have used the Kepler Quarter-9 (Q9) long-cadence (30 minute sampling) light curves, as processed by the Kepler pipeline, to characterize the photometric variability of the sample set of stars on timescales comparable to the transit durations (0.25 days).  The Q9 light curves span 98 days (beginning at BJD = 2454808), and represent one of the best Kepler quarters in terms of the smallest loss of data due to spacecraft anomalies or external features such as solar flares (see Kepler Data Release 21 notes).

For each Kepler candidate exoplanet host star in our sample, we calculated an average light curve rms dispersion.  We set a 0.25 day ``window'' around each data point in the light curve and calculating the rms dispersion within that light curve for an 0.25 day window, the number of light curve data points per window was 12 (assuming no 30 minute cadences were missing in the light curve).  The 0.25 day windows typically started with the 6th data point and ended with N-6th data as the light curve data points near the light curve end would not span 0.125 days in both directions.  For a typical light curve of approximately 4650 data points, approximately 387 rms dispersions per light curve were calculated and a final average 0.25-day rms dispersion was computed.  The photometric noise contribution to the rms dispersion was estimated from the 0.25-day dispersion calculated for all 4600 KOIs by fitting a function to the floor of the photometric dispersion as a function of the stellar Kepler magnitude (e.g., Gilliland et al. 2011, Ciardi et al. 2011).  The magnitude dependent photometric noise estimations are subtracted from the rms dispersions in quadrature, leaving only an estimate of the intrinsic (instrumental plus astrophysical) light curve dispersion.  The photon-noise floor is very similar to the noise floors reported previously (e.g., Gilliland et al. 2011, and Ciardi et al. 2011).

We have chosen to concentrate on the 0.25 day time bin as this duration is fairly typical habitable zone exoplanet transit ingress, egress, and duration.  Ciardi et al. (2011) provided a similar light curve analysis but for all Kepler stars (not just small exoplanet hosts) and that work shows the general variability levels present (Fig. 8 in Ciardi et al. 2011). Their results show for all stars, and we have confirmed this as well for our sample of small exoplanet host stars, that time bins from 0.2 to 0.5-day provide similar results.  A direct comparison with Ciardi et al. (2011) is not optimal, as that work used early release Quarter-1 data which contained some processing problems, albeit mainly related to long term trends (e.g., Kepler Data Release 21 notes; McQuillan, Aigrain, \& Roberts 2012). Further, we have chosen to use the rms dispersion, instead of the Kepler-specific combined differential photometric precision (CDPP), as the rms dispersion is an easily calculable quantity with a direct statistic connection to other well-known statistics (e.g., mean absolute deviation).  In general, the rms dispersion and the CDPP are highly correlated and nearly linear in relationship with a slope of $\sim 3$ for the 6-hour timescale.  The sample of small exoplanet hosts stars presented here lie near the bottom of the Ciardi et al. (2011) distribution of rms dispersion vs magnitude.

Our star sample harbors planetary candidates of radii less than 2.5 $R_\oplus$.  Figure 2 plots the 0.25-day rms dispersion (in milli-magnitudes) of the stars as a function of the transit depth (top) and the derived planetary radii (bottom), as listed in the Q1-Q12 catalogue paper (Rowe et al. 2015).  
The transit depths and planet radii plotted represent those of the
smallest transiting planet in the system in cases where more
than one planet is detected. 
The errors in the determined planet radii are dominated by the uncertainties of the stellar radii.  If the stellar radii were known perfectly, the typical planet radius would be measured (i.e., the transit depth) to better than 0.5\%.  At fainter magnitudes ($Kp \gtrsim 14.5$ mag), the light curve scatter is dominated by the photon-noise and even with subtraction of the noise-floor, there remains an uncertainty in the stellar variability properties as a result of the larger photon-noise.  We estimate that for the brighter stars ($Kp \lesssim 14.5$ mag), the remaining dispersion (after the subtraction of the floor) is $\lesssim 50$ ppm, but for the fainter stars ($Kp \gtrsim 14.5$ mag), the remaining dispersion is $\gtrsim 150 - 250$ ppm.  Variability beyond these levels may be a result of stellar variability.

In the top panel of Figure 2, we note that for transit depths of $\sim$0.2 mmag and larger, the distribution with light curve variability appears flat while for the smallest transit depths, the most quiet stars are required for transit detection. The trend of smaller planets detected about quieter stars (bottom panel) is less obvious (due to observational factors such as short period orbits providing multiple transits), but a turnover near 1 to 1.2  Earth radii is indicated. The significance of the smaller light curve standard deviations for stars harboring planets with smaller transit depths and radii was tested using the Kolmogorov-Smirnov (K-S) test.  To do this, the sample was divided into two, placing those with smaller than a given transit depth or planet radius in one subsample and the remainder in the other. In Figure 3 we show the statistical significance for the choice of the value about which the subsamples are defined.  Here, $P_{KS}$ is the probability that random fluctuations alone could be responsible for the differences between the two distributions. As seen in both panels, the trends we see in Figure~2 are statistically significant. 

To test the robustness of this significance, we perform a set of Monte
Carlo simulations, re-using the data of Figure 2, but giving each
point a random fluctuation in transit depth or planet radius according
to the uncertainties in each property.  The transit depth
uncertainties are taken from the NExSci Cumulative list of KOI
properties and the uncertainties in planet radius are adopted from
isochrone fits using the spectroscopic stellar properties of Everett
et al (2013).  Figure 4 shows the results of $10^4$ simulations.  In
panel a, the critical transit depth is found to be $0.18\pm0.02$~mmag.
Panel b shows the distribution of its significance over all tests (all
tests show significantly quieter light curves among stars with the
shallowest transits).  Panel c shows the critical planet radius
($1.16\pm0.23$\re) that best divides the sample into two and
panel d the corresponding distribution of its significance (most
simulations show significantly quieter stars among those with the
smallest transiting planets at the 99\% confidence level).

That is, transit depths of less than $\sim$0.18 mmag, ($\la$1.25 \re~planets if orbiting an assumed sun-like star) detected in the $Kepler$ data are quite rare, as these shallow depths require the most photometrically quiet stars (transit timescale sigma values $\le$0.1 mmag) - even after the photometric noise has been accounted for. This is not to say that only quiet stars have earth-size analogs, but rather $Kepler$ can only detect earth-size analogs around the quietest stars ($\sigma_{0.25day} \lesssim 0.5 $mmag).  These planet candidates were found in the first 12 Quarters of $Kepler$ observations; yet, the mission ran for four years (16+ Quarters), long enough to hopefully detect an additional 1 or 2 transits of small, habitable zone planets orbiting solar-like stars. As the completeness of the Kepler pipeline is tested and better understood (e.g., Christensen et al. 2015), it will be interesting to understand how the extra year of data improves the situation.  

\subsection{Subgiant Variability}

In order to gain insight on the sources of ``stellar noise" we first consider some results from radial velocity studies and their possible relationship to photometric variability.
Radial velocity measurements of exoplanet host stars (e.g., Hartman et al., 2011) 
have reported an issue of RV jitter in their spectroscopic measurements when observing evolved F and G (subgiants).
Saar et al. (1998) suggested that jitter seen in the measured position of spectral lines is caused by convective
inhomogeneities that vary with time rather than thermal inhomogeneities such as star spots.
Jitter values for F dwarf stars can range from a few up to 30-50 m/sec and 
cause a varying RV signal which adds a systematic
noise to formal velocity solutions such as those used to detect exoplanets.
Since RV jitter is observed in the hotter solar-like stars (early G and F subgiants and evolved 
main sequence stars) and is apparently due to large scale
convective surface phenomena, we might also expect to see larger photometric variability in such stars 
\footnote{
We note here that the photometric ``flicker" variations discussed by Bastien et al., (2013) occur
on minute to hour timescales and are assumed to be caused by granulation noise. 
Their result relating the ``flicker" value to a star's log g works well for giants and some
hotter subgiant stars for which the granulation noise happens to modulate on $\sim$8 hour timescales - 
a property of these star's atmospheres. Kallinger et al. (2014) show that the ``flicker" metric as 
formulated in Bastien et al. does not provide an accurate measure of log g 
for early subgiants and main sequence stars.}.

To understand how a relationship between RV jitter and photometric variability might work,
let us look at the likely photometric variability
signature to be introduced into a light curve due to a typical spectroscopic RV jitter. 
Kepler-21, a F5 IV star, is a good example of an exoplanet host star with small photometric variability
but exhibiting RV jitter. 
Howell et al. (2012) confirmed a 1.6 \re~planet in this system orbiting every 2.8 days about the star. 
However, attempts to estimate the mass of this exoplanet, Kepler-21b, via radial
velocity measurements made with Keck HIRES were
unsuccessful largely due to RV jitter, yielding only an uninteresting upper mass limit of 10 $\me$. 
During the course of the study of Kepler-21, forty high-resolution 
spectra were obtained over various timescales
ranging from a few minutes to approximately 100 days. The RV jitter measured in this subgiant host star was
near 7 m/sec which translates into (using the equation presented in Ciardi et al. 2011) 
$\sim$0.15 mmag (130 ppm) of photometric variation that would be present in Kepler-21's light curve. 
Statistical analysis of the Kepler-21 Quarter 9 light curve for a 0.25 day binning
yields a light curve $\sigma$=0.13 mmag, a value which, by the way, remains relatively constant for this star over many quarters of $Kepler$ observations. 
Thus, Kepler-21's light curve standard deviation (0.13 mmag)
is comparable to the photometric noise expected to be
introduced into the light curve by radial velocity jitter (0.15 mmag). 
Examination of Figure 2 shows 
that the Kepler-21 level of 0.25 day photometric variation, 1.3$\times$10$^{-4}$ (0.13 mmag), 
is near the smallest
light curve dispersions measured for any of the fainter stars in our entire sample.
The relatively large RV jitter observed in this star does not seem to correlate with large photometric variability, therefore the transit method is more sensitive to finding planets in such stars, especially the smallest planets.

Given that significant RV jitter may produce global photometric variations, 
possibly limiting the detections of the smallest exoplanets, 
we examined the level of variability in our light curve sample with respect to the location of the stars 
in a pseudo H-R diagram. If RV jitter increases for hotter stars or 
for stars moving into the subgiant region,
we would expect to see a trend toward increasing light curve variability as the stars evolve. 
Figure 5 examines such an assertion plotting again a pseudo-HR diagram but with the stars now color coded
by their 0.25 day variability level. This figure reveals that as the stars evolve away
from the main sequence and are presumed to have
more RV jitter (as evidenced by spectroscopic observations, e.g.,
Hartmann et al., 2011; Marcy et al., 2014), 
they do not show a similar increasing trend in their photometric variability.
We see that the stars act as individuals, that is photometric variability in 0.25-day bins does not seem to 
follow any discernible pattern.

\subsection{Transit Detectability}

Stellar variability can easily overwhelm the subtle signature of a transit by a
terrestrial-size exoplanet.  Therefore, it is not surprising that photometric quietness
is a distinguishing property in our sample of detected $Kepler$ hosts of small
exoplanets. A relevant astrophysical question--that is also of importance in devising
future exoplanet search strategies--is whether the photometric quiescence in our
ensemble is the result of a bias arising from an exoplanet-selected sample or if it
reflects an underlying stellar property.  The emerging results from Bastien et al.
(2015) are particularly relevant in this regard.  These investigators
conducted a Ca {\small II} H and K survey of ostensibly solar-type stars in the $Kepler$
field without regard to exoplanet detection in their sample, and have compared their Ca
{\small II} measures with the range of photometric variability exhibited in the
$Kepler$ light curves.  Bastien et al. find that 93\% of the 167 objects that exhibit low
photometric variability, i.e., similar to that of the average Sun or even lower, are in
fact subgiants, as based on an analysis of the flicker properties of the light curves
and the correlation of flicker with surface gravity (Bastien et al. 2013).  In their
analysis, subgiants are classified as those stars with log $g <$ 4.2, corresponding to
about 40 stars in our sample and
nearly 100\% of their photometrically quiet stars have
log $g <$4.4, which contains the majority of our sample (see Figs. 1 \& 5). Thus, while subgiants
may generally be photometrically quiet, Figure 5 shows that not all of them are equal in
their variability.

Though not directly related to photometric variability in the visible band, this finding
is reminiscent of earlier work by Wright (2004) who applied $Hipparcos$ parallaxes to
the most chromospherically quiescent stars in the Mt. Wilson survey, that is, stars
originally identified by Baliunas \& Jastrow (1990) as solar analogs but with
chromospheric activity that is even lower than solar minimum values and, therefore,
could be Maunder Minimum candidates.  However, Wright (2004) determined that these
objects were not solar analogs but are actually more evolved, older (subgiant) stars.
This conclusion, while suggestive, does not by itself mean that these chromospherically
quiescent objects must be photometrically quiet in the visible band.  However, Johnson
et al. (2011a,b) find results of direct relevance to the context of our work from their
Keck spectroscopic survey of subgiant stars in search of Jovian companions combined with
a parallel photometric monitoring program.  In particular, Johnson et al. (2011a) find
in their monitoring of two subgiants with Jovian-size planets that both stars were
constant in brightness to $\leq$ 2 mmag.  In an expanded study of 18 subgiants
spanning a broad range of physical properties, Johnson et al. (2011b) do not find any
evidence for photometric variability in their sample to within their limits of
precision, which were typically in the range of $\sim$ 3-6 mmag.  While these
variability studies have been limited in sample size and duration, the results of this
work and the aforementioned investigations are suggestive that photometric quiescence in
the visible band is a general property of subgiant atmospheres.  

The detectability of a transit is enhanced if the contribution by stellar noise is
minimized, which appears to be the case in stars somewhat more evolved than the Sun.
Conversely, the transit amplitude is reduced for subgiants relative to dwarf host
stars at a given exoplanet size. In addition to photometric quiescence and signal
quality (related to the apparent brightness of the star), orbital parameters and
geometric constraints govern the detectability of exoplanet transits.  These factors
favor exoplanet systems in proximity to a large star since the allowable range of
orbital inclinations for the occurrence of a transit within our line-of-sight is
larger and the frequency of transits is greater.  Therefore, our sample could reflect
a bias toward compact exoplanet systems with evolved host stars if the transit
frequency is high.  However, if the average transit frequency is closer to $\la$1 per
quarter of $Kepler$ data, then other exoplanet system architectures merit
further quantitative examination.

As discussed by Giampapa et al. (1995), a transit can be seen from Earth only if the
inclination of the orbital plane of an exoplanet is within a small angle,
$\Delta\theta$ = $R_{s}/a$, of our line of sight, where $R_s$ is the stellar radius
and $a$ is the semi-major axis of the exoplanet orbit.  The probability of observing
a transit at a given time also depends on the ratio of the transit duration to the
orbital period, which is proportional to $R_{s}/a$.  Therefore, the number of
exoplanet systems with a potentially observable transit is proportional to
$(R_{s}/a)^2$.  Of course, the transit signal depth varies as the inverse square of
the stellar size, i.e., as $(R_{p}/R_{s})^2$, where $R_p$ is the exoplanet radius.
In order to gain insight on these competing parameters for transit detectability, we
determine at what stellar radii these dimensionless numbers become of comparable
importance, or

$$
R_{s}~=~M_{s}^{1/6}~P^{1/3}~{R_{p}}^{1/2}~~,
$$

\noindent{where $R_s$ is in solar radii, $M_s$ is the stellar mass in solar masses, $P$ is the period in years and $R_p$ is the exoplanet radius in earth radii.  
In equation (1) we applied Kepler's Third Law to
convert from semi-major axis to the observable of orbital period.} 

The relation in (1) states that, for a given exoplanet radius, the stellar size
increases with orbital period in order to present a larger cross section for a
transit to be visible within $\Delta\theta$ of our line-of-sight.  At the same time,
at a given orbital period the host star radius increases with increasing exoplanet
size in order to yield a transit depth at a fixed amplitude.  We plot in Figure 6 the
stellar radius as a function of exoplanet orbital period as calculated from equation
(1) for exoplanet sizes of 1.0, 1.5 and 2.5 earth radii, respectively.
We note, parenthetically, that since the relation in (1) is only weakly dependent on stellar mass and the mass range of our sample is narrowly centered on a solar mass, we neglect the leading term in equation (1).  The curves in Figure 6
essentially divide the $R_s$--$P$ plane into two detection regimes: above a curve for
a given exoplanet radius the transit depth declines (since the host star is larger)
while below each curve the geometric constraint for the visibility of a transit in
our line of sight becomes more restrictive (since the host star is smaller).
Inspection of Figure 6 reveals that, at orbital periods in the broad range of roughly
$\sim$ 0.3 years to $\sim$ 1 year, host star sizes in the subgiant region are
indicated for exoplanet radii greater than one earth radius.  This range of orbital
periods is still consistent with the possibility of detecting multiple transits
occurring in the first 12 Quarters of the $Kepler$ mission.  Thus, in brief summary,
the stellar characteristics of our sample appear to be the result of the combined
factors of the intrinsic stellar properties of photometric quiescence and stellar
size, exoplanet system architectures that range from in-proximity to $\sim$ 1 AU of
the host star, and extrinsic characteristics such as apparent brightness that lead to
a detection bias toward the subgiant regime, as reflected in Figures 1 \& 5. 


\section{Photometric Variability and Activity in Solar-Type Stars}

As noted above, an understanding of the nature of intrinsic stellar variability, 
as seen in both
broad photometric bands and spectral lines, can yield insight on the origin of detection biases in searches
for terrestrial-size planets around solar-type stars.  The next step beyond detection is exoplanet system
characterization, particularly in the context of habitability.  In this regard, stellar
variability manifests itself as a modulator of the radiative flux and energetic particle 
environments in which exoplanet
atmospheres form and evolve.  In view of these considerations, we examine the relationship between broad-band
photometric variations and magnetic field-related activity in sun-like stars and the Sun itself. We also
discuss the role of intensity fluctuations due to granulation noise as an additional non-magnetic (i.e.,
star-spot) component of photometric variability.

\subsection{Solar-type Stars and the Sun}

Prior to the $Kepler$ mission, the most extensive, long-term study of the joint behavior of mean chromospheric
emission and brightness changes in solar-type stars utilizing high-precision, ground-based differential
photometry is summarized by Lockwood et al. (2007).  Hall et al. (2009) discuss an extension of this effort to
a larger sample of more nearly sun-like stars.

In both the Lockwood et al. and the Hall et al. investigations, a clear correlation between increasing rms
variation in the Str\"{o}mgren photometric $b$ and $y$ bands, and the logarithm of the mean level of
normalized chromospheric Ca II emission, emerges at chromospheric emission levels greater than that of the Sun
(by $\sim$ 0.1 -- 0.2 dex).  However, as concluded by Hall et al., brightness variations and changes in
activity at solar-like levels appear to be uncorrelated in the nearby bright solar analogs, including the
solar twin, 18 Sco.  We can infer from these studies, and the case of the Sun (see below), that the logarithm
of the rms of the photometric variation is relatively low ($\sim$ -3.0 -- -3.5) at solar-like values of
chromospheric emission in sun-like stars.  Therefore, solar-like host stars with chromospheric emission levels
similar to the quiescent Sun likely will be characterized by uncorrelated but solar-like (or lower) amplitudes
of brightness variations in the visible ($Kepler$) band.  


A comparison of the stellar results with the Sun-as-a-star in this context becomes  appropriate given that (1)
the amplitude of the solar irradiance variability in the visible at the 1 -- 2 mmag level is similar to that
in our $Kepler$ sample, (2) near-simultaneous, superb space- and ground-based data are available for the
Sun-as-a-star, and (3) the Sun is the host star to a planetary 
system that includes terrestrial-size planets.  The solar
measurements uniquely provide time series of both spectroscopic and broadband data that can be compared with the
results from the $Kepler$ data of nearly continuous photometry but limited, simultaneous spectroscopic data.
The time series of chromospheric Ca {\sc II} K line (hereafter referred to as K-line) strength recorded for
the Sun as a star, as modulated by the solar cycle of activity, can then be compared with photometric solar
data in other bandpasses  to gain insight on the joint response to activity variations analogous to what may
occur in sun-like stars. 

We utilize the time series of K-line spectra obtained by the Integrated Sunlight Spectrometer (ISS), which is
one of a suite of high-precision instruments that comprise the {\it Synoptic Optical Long-term Investigations
of the Sun} (SOLIS) facility of the National Solar Observatory on Kitt Peak, Arizona. Keller, Harvey \&
Giampapa (2003) give a description of the SOLIS instruments.  The {\it Solar Radiation and Climate Experiment}
(SORCE) satellite is the source of solar irradiance data in selected bands.

High-resolution (R $\sim$ 300,000) spectra centered at the K-line have been obtained on a daily basis (weather
conditions and instrument status permitting) by the SOLIS ISS since inception in December 2006. We adopt the
parameter time series for the 1 {\AA} (0.1 nm) bandpass centered at the K-line at 3933.68 {\AA}, which is a
standard data product produced by the SOLIS program for the community (http://solis.nso.edu/iss). A typical
measurement error in this parameter is $\sim$ 0.001\%. Note that this bandpass is proportional to the Mt.
Wilson S-index, hence, it can be readily compared to stellar data via calibration relations (e.g., Hall \&
Lockwood 1995).  

The ISS K-line 1 {\AA} time series begins in Figure 7 in the declining phase of Cycle 23, extending through the
protracted 2008-2010 solar minimum and continuing through the current rise toward maximum in Cycle 24.  Note,
parenthetically, that Cycle 24 thus far appears to be weaker in amplitude than the previous Cycle 23 that, in
turn, had a peak sunspot number that was roughly a factor of two below the strong maximum of Cycle 19 in
the 1950s. From a historical perspective on the sunspot record of the past 400 years, the current Cycle 24
would be considered 'moderate' in its Sunspot Number Index -- exceeding that of Cycles 5 and 6 that
characterized the Dalton minimum during the early 1800s though clearly lower in amplitude than the peak of the
solar cycles observed in the modern era (see Clette et al. 2014, their Fig. 65).  Therefore, the ISS data are
more representative of a quiescent Sun-as-a-star, similar to the $Kepler$ stars in this study. 

We compare the relative strength of the K-line core with the broad band flux emitted by the Sun detected at 1
AU by the SORCE satellite, calibrated in absolute monochromatic flux. Instruments on board SORCE record total
and spectral irradiance data for the Sun extending from the X-ray (0.1 nm) through the near infrared at about
2.7 $\mu$m with a variable spectral resolution of 1 nm -- 34 nm over the entire spectral range (Woods et al.
2000).  The SORCE Spectral Irradiance Monitor (SIM) instrument yielded monochromatic absolute flux
measurements in the 310 nm -- 2400 nm bandpass from 2003 April to 2011 May (see Rottman, Woods \& McClintock
2006 for a description of the SIM instrument).  The accuracy of the daily monochromatic flux measurements is
2\% while the errors in precision (i.e., long-term repeatability) are $<$ 0.1\% per year.  We utilized those
SIM data that overlapped with the $Kepler$ visible bandpass of approximately 400 nm -- 900 nm. 

Selecting SIM and SOLIS/ISS K line data obtained for the same Julian Day number yields the scatter plot
displayed in Figure 8. It is evident by visual inspection of Fig. 8 that there is no or very little
correlation of the relative flux in the core of the K-line with the solar flux measured by the SIM instrument
in the $Kepler$ visible bandpass.  Hence, it is probable that the flux from solar-like stars in the $Kepler$ 
bandpass is effectively independent
of any chromospheric activity in the stars when at primarily quiescent solar-cycle levels.

These results for the Sun and for solar-type stars suggest that the selection of quiet stars for the purposes of
achieving high-sensitivity limits for the detection of photometric transits of earth-size planets cannot be
guided by chromospheric emission levels alone: activity can be an ambiguous guide to the predicted amplitude
of variability in the stellar light curve at solar-like activity levels (or even at levels a factor of two or more
greater, see Bastien et al. (2013)) 
even after taking into account possible inclination effects with respect to the line-of-sight (Hall et al.
2009).   Similarly, low-amplitude variability in the stellar light-curve is not necessarily an indication of
quiet chromospheric activity, which is particularly relevant to the selection of stellar samples for
measurements at the highest possible Doppler precisions.  For example, inclination effects can lead to the
observation of low-amplitude, broad-band variability in stars with otherwise high chromospheric and coronal
emission.

\subsection{Photometric Noise Due to Granulation in Solar-type Stars}

Trends in granulation noise, or "flicker" (Bastien et al. 2013), with stellar type due to contrast
fluctuations merit discussion since the timescale of this form of intrinsic stellar variability is relevant
to photometric transit timescales.  According to model simulations of surface convection in late-type, main
sequence stars (Beeck et al. 2013a,b) and a limited grid of late-type main-sequence and giant stars (Trampedach
et al. 2013), granulation cells exhibit higher rms intensity fluctuations relative to the time-averaged mean
granular intensity, and they become larger in size, toward higher effective temperatures and lower gravities
(i.e., movement toward luminosity class IV).
Therefore, we might expect a relatively lower number of small exoplanet 
detections toward the upper left in Fig. 1 because of
increased granulation contrast fluctuations, i.e., granulation noise, that adds to the already present transit signal bias against the detection of terrestrial-size planets around larger (and lower gravity) stars.

At cooler temperatures and higher gravities, the size of granulation cells decreases (by a factor of $\sim$ 25
from early F V to early M V) and the contrast fluctuations are reduced.  Therefore, we would expect to see
more detections in the G -- K range where the granulation noise is decreasing and the stars are still
relatively bright.  Granulation cell size and noise decreases toward M stars where the convective energies and
velocities are lower, and the brightness substructure is reduced (Beeck et al. 2013a,b). However, M stars are
also fainter so we begin to lose observational sensitivity in this regime. The
paucity of detections along the ZAMS may be due to (a) not many young stars in the $Kepler$ field and (b)
increased activity in younger stars that adds to the granulation noise.
By contrast, Fig. 5 does not exhibit the expected trend in stellar variability due to granulation noise alone
as suggested by the results of the above model simulations of surface convection along the main sequence.
Therefore, (a) the models may not be correct, (b) the models are not applicable to the 0.25-day timescale
adopted in Fig. 5, or (c) other photometric noise sources dominate any intensity fluctuations due to
granulation on this timescale (see Kallinger et al., 2014).  

We note that the overall trends in granulation with fundamental stellar properties, as deduced from the above
simulations,  appear consistent with the observed variations in the amplitude of photometric jitter with
stellar effective temperature and gravity.  Cranmer et al. (2014, their Fig. 4) find that the amplitude of
$Kepler$ photometric light curve flicker is at a minimum in the range of 
surface gravities and effective temperatures
represented by our host star sample, particularly as seen in their empirical model that includes magnetic
suppression effects on granulation.  While magnetic effects would seem to imply an enhanced stellar noise
contribution to photometric jitter due to magnetic activity, the results for solar-type stars and the Sun as
discussed above demonstrate that magnetic activity and photometric variability are uncorrelated in the
$Kepler$ visible band, at least for quiet chromosphere stars.

\subsection{Characterizing Activity at UV Wavelengths in $Kepler$ Host Stars} 

Obtaining information on emission levels in the ultraviolet, especially the far ultraviolet where the spectrum
of solar-type stars can be dominated by emission lines from activity, yields critical input for studies of the
structure and photochemistry of exoplanet atmospheres with implications for astrobiology and the detection of
biosignatures (e.g., Canuto et al. 1983; France et al. 2013).  Smith and Redenbaugh (2010) found that the
magnitudes of solar-type stars in the FUV bandpass of the $GALEX$ satellite were correlated with chromospheric
Ca {\small II} emission strength for a sample of field stars drawn primarily from the Mt. Wilson HK Survey
(Vaughan et al. 1978; Baliunas et al. 1995), at least for levels of normalized chromospheric Ca II emission
that exceed the level of the mean Sun (Linsky et al. 1979). Unfortunately, there is a paucity of measured FUV
magnitudes available from the $GALEX$ all sky survey for our relatively fainter sample of $Kepler$ host stars
nor are Ca {\small II} H and K data available yet.  In anticipation of the latter soon becoming available, we
briefly examine the solar data to assess K-line core emission as a reliable predictor of the level of FUV
emission fluxes in sun-like stars.

In order to do so, we utilize the SORCE data in the far ultraviolet band from 115 nm - 180 nm, as obtained
with the {\it Solar Stellar Irradiance Comparison Experiment} (SOLSTICE; Rottman et al. 2006). This band
includes the strongest emission line feature in the FUV, namely, Ly$\alpha$, that also could have an important
effect on exoplanet atmospheric structure (Knutson et al. 2010; Linsky et al. 2014).  The accuracy of the
SOLSTICE monochromatic flux measurements is in the range of 1.2\% -- 6\% while long-term repeatability is
0.2\% -- 0.5\% per year. The overlap of the SOLSTICE data with the SOLIS ISS data extends to 15 July 2013 or
JD 2456489 (see Fig. 7). The clear correlation between the solar far UV flux and the chromospheric K-line
relative strength in Fig. 9 is in striking contrast to the lack of correlation between the K-line and the
$Kepler$ bandpass in Fig. 8. 

The correlation in Fig. 9 essentially confirms that the solar chromospheric K-line core emission and the
total flux in the far UV bandpass, respectively, have qualitatively similar origins:  each is dominated by
emission from magnetic active regions.  By contrast, the $Kepler$ optical  bandpass is obviously dominated by
photospheric emission. Thus, the variations in the $Kepler$ bandpass are much smaller in relative amplitude
than in either chromospheric spectral lines or broad photometric bandpasses that include a significant
radiative cooling component resulting from magnetic field-related, non-radiative heating, such as the far UV
or X-ray  bandpasses observed by the SORCE satellite.  Thus, the solar data suggest that Ca {\small II}
resonance line emission can be used to estimate the level of far ultraviolet emission present in sun-like
stars. This is somewhat at variance with the results of Smith and Redenbaugh (2010) who found the onset of a
correlation with their ultraviolet color excess (i.e., a measure of relative FUV emission due entirely to
magnetic activity) only at levels of normalized chromospheric Ca {\small II} emission in their stellar data
that were a factor of $\sim$ 2 higher than the same index for the average Sun. In other words, their estimated
ultraviolet color excess  appears uncorrelated with chromospheric emission levels at or below that of the
average Sun. In anticipation of K-line spectral data eventually becoming available for many $Kepler$ host stars,
we give the results of a linear regression of the Sun-as-a-star data in Fig. 9, or

\vspace{0.1in}

$
{\rm log~(FUV)}~=~1.419(\pm 0.042)~$\rm log (K)$ ~ -~(0.181\pm 0.040) ~ ,
$

\vspace{0.1in}

\noindent
where FUV is in W-m$^{-2}$, the K parameter is the 1 {\AA} K-line index centered at 3933.68 {\AA}, and the
formal 1-$\sigma$ errors in the fitting coefficients are given in parentheses.  The range of applicability of
this relation is given by the range shown in Fig. 9.  In brief summary, the example of the Sun would seem to
affirm that K-line core emission is a reliable predictor of the level of far ultraviolet emission in sun-like
stars.  The apparent disagreement with the stellar results of Smith \& Redenbaugh (2010) at sun-like levels of
chromospheric activity could be due to the strong sensitivity to errors in the correction for the photospheric
contribution to an activity diagnostic at low levels of activity.  However, this suggestion will require
investigation beyond the scope of this paper.

\section{Summary}
The spectroscopically vetted Everett et al. (2013) 
sample of over 200 small ($\le$2.5 R$_{Earth}$) exoplanet host stars in the $Kepler$ field has provided 
the ability to measure their photometric variability on transit timescales of interest and relate it to
their stellar properties.
We note that detections of small, $<$2.5 \re~exoplanets, those with photometrically small transit depths, 
are preferentially detected among the quietest $Kepler$ stars. 
We examine simultaneous spectral and photometric 
observations of our Sun, use these as a proxy for solar-like stars, 
and relate these data to the general solar-like stars in this study.
$Kepler$ observations are shown to be insensitive to detecting variability due to chromospheric activity.
We note for the Sun, and postulate for the $Kepler$ sample, that photometric quietness in the optical bandpass of
$Kepler$ does not translate directly into an inactive star with high radial velocity stability.
Likewise, RV jitter common in brighter or more evolved stars, a property that often disqualifies 
these stars for planet searches by the Doppler technique, does not mean that they are photometrically noisy.
In fact, some of these stars observed by $Kepler$ represent the photometrically quietest stars in the sample.
While the trend that small exoplanets are more easily detected orbiting photometrically 
quiet stars is a known observational bias that must be accounted for when exoplanet occurrence rates are
estimated, we find that transit depths less than 0.17 mmag, roughly corresponding to 1.25 \re~ planets, are quite
rare detections in $Kepler$ observations as they require bright, very photometrically quiet stars. 

\acknowledgements{
We wish to thank the staff of the $Kepler$ project, the NASA Exoplanet Archive, and the 
Kitt Peak National Observatory
for their continued support of the $Kepler$ mission and its data products. 
MEE wishes to acknowledge funding supporting this work at NOAO provided by NASA agreement number NNX13AB60A. 
}

{\it Facilities:} Kitt Peak National Observatory, Mayall(RCSPEC), $Kepler$, National Solar Observatory, SOLIS(ISS),
NASA Exoplanet Archive

\clearpage

\begin{figure}
\includegraphics[angle=-90,scale=0.7,keepaspectratio=true]{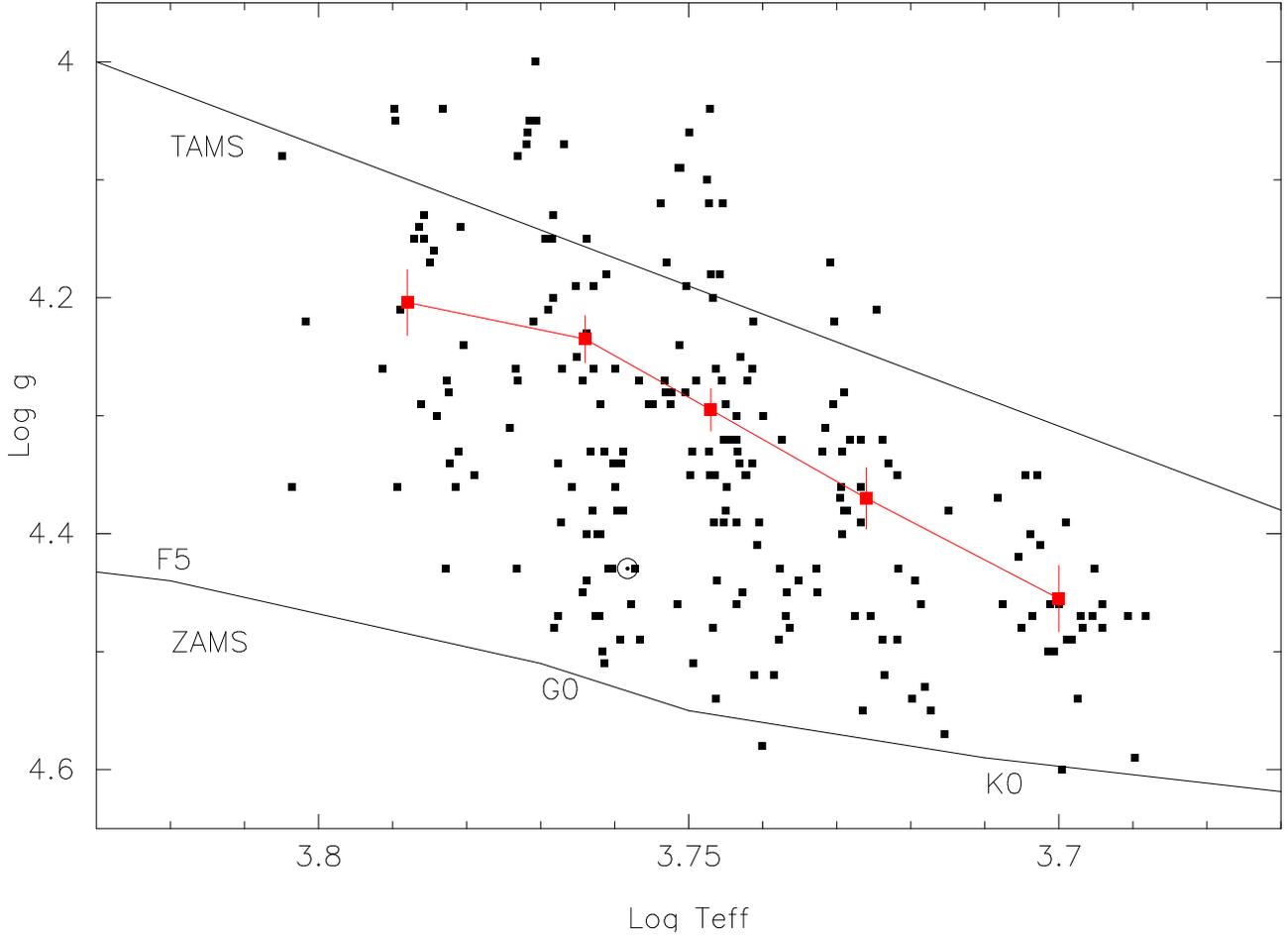}
\caption{The $Kepler$ exoplanet host star sample of Everett et al. (2013) 
discussed in this paper plotted in the Log 
\teff-log g plane. 
The solid lines mark the location of the Zero-Age Main Sequence (ZAMS) and Terminal Age Main Sequence (TAMS) 
limits (adapted 
from Schaerer et al. 1993). The $\odot$ shows the Sun's location
and the main sequence spectral types F5 to K0 are marked. 
The formal 
adopted (1$\sigma$ uncertainties) for the points are 75 K for \teff~ and 0.15 for log g (Everett et al., 2013). 
The red points show the relative density of occurrence for the stars binned into five temperature regimes 
with uncertainties based on root N counting statistics.
Note the evolutionary spread away from the main sequence.
}
\end{figure}

\begin{figure}
\includegraphics[angle=-90,scale=0.7,keepaspectratio=true]{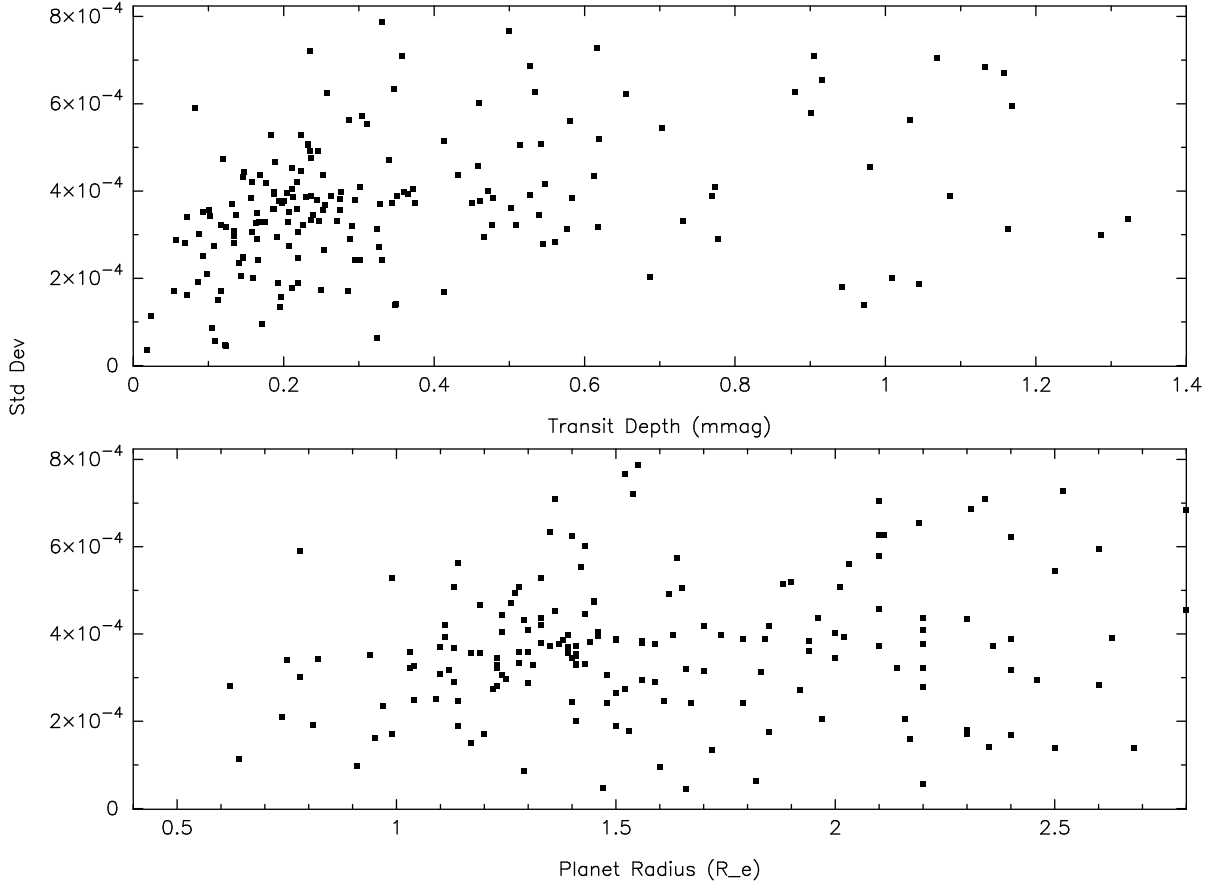}
\caption{(Top) The transit depth (mmag) measured from $Kepler$ light curves for planets orbiting our sample stars 
is plotted against their 0.25 day (6 hr) light curve standard deviation (mag) as described in the text.
A typical transit depth uncertainty, as reported in the NASA Exoplanet public archive, is $\pm$6$\times$10$^{-5}$
mag.
(Bottom) The derived
radius of the smallest planet orbiting each sample star (in Earth radii) vs. the 
0.25 day (6 hr.) light curve standard deviation in magnitudes. 
We see that, in general, the smallest detected transits require the photometrically quietest 
stars and that a change in detection efficiency occurs near transit depths of 0.17  mmag or
1.25 \re, where only the quietest stars are represented (see Figure 3). 
Note, an Earth-size planet orbiting a Sun-like star has a transit 
depth near 0.1 mmag ($\sim$100 ppm).
}
\end{figure}

\begin{figure}
\includegraphics[angle=-90,scale=0.7,keepaspectratio=true]{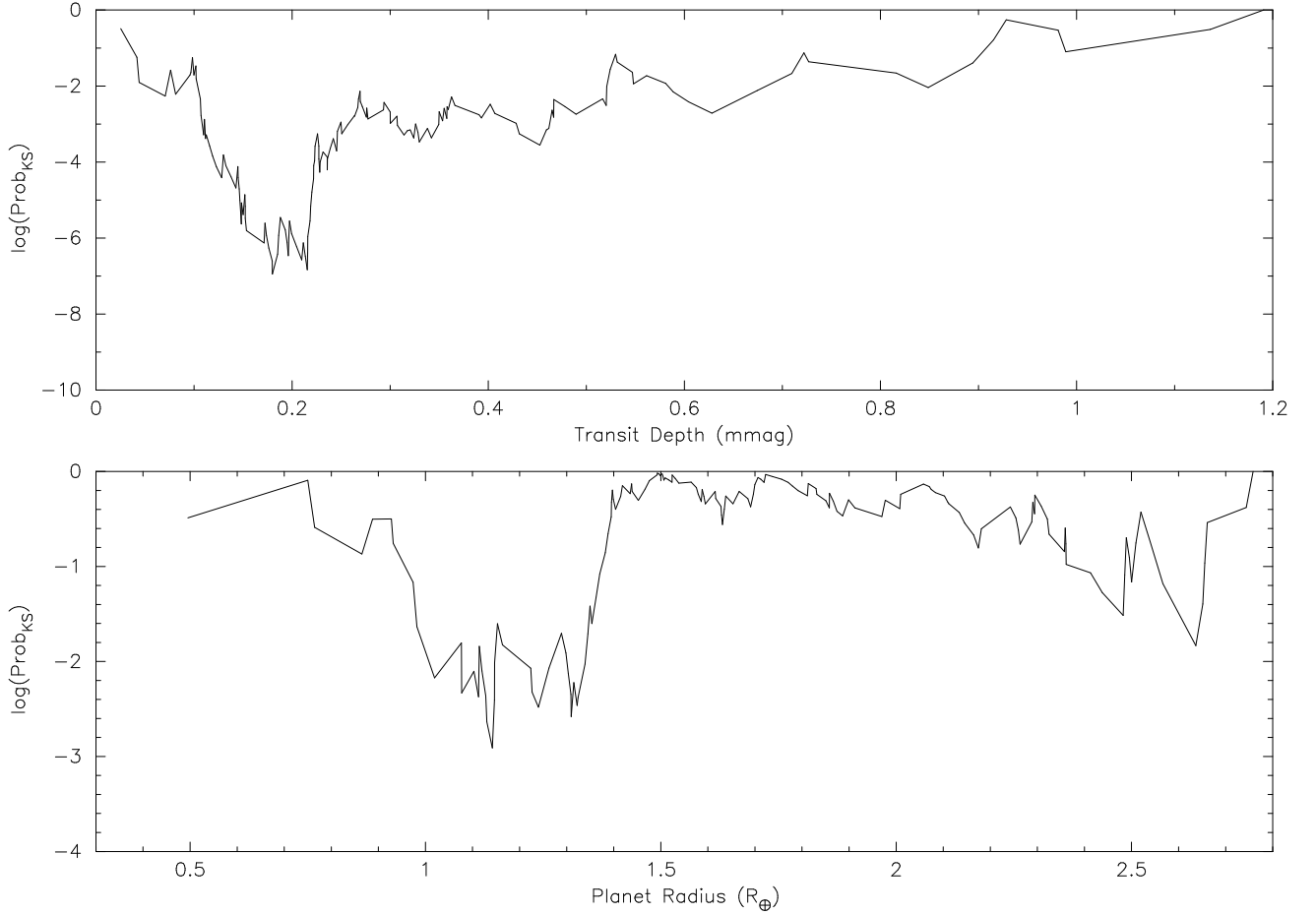}
\caption{
(Top) The significance of the difference between the standard
deviations found in 0.25~day (6~hr) light curves for stars with
transits shallower than a given transit depth compared to those stars
with deeper transits is plotted as a function of the given transit
depth. $P_{KS}$ is the probability, based on the K-S test, that random
fluctuations alone are responsible for the relatively quiet stars
found in the sample with shallow transits (as seen in Figure 2). The
curve has a deep minimum near 0.18 mmag, indicating a significant
threshold transit depth. (Bottom) Similar to the top panel, the light
curves are split across a given planet radius. Once again, the
appearance of Figure 2, where relatively quiet stars are favored among
the sample of smallest planets, is significant. The most significant
dividing point in planet radius occurs near a value of 1.2\re.
}
\end{figure}

\begin{figure}
\includegraphics[angle=-90,scale=0.7,keepaspectratio=true]{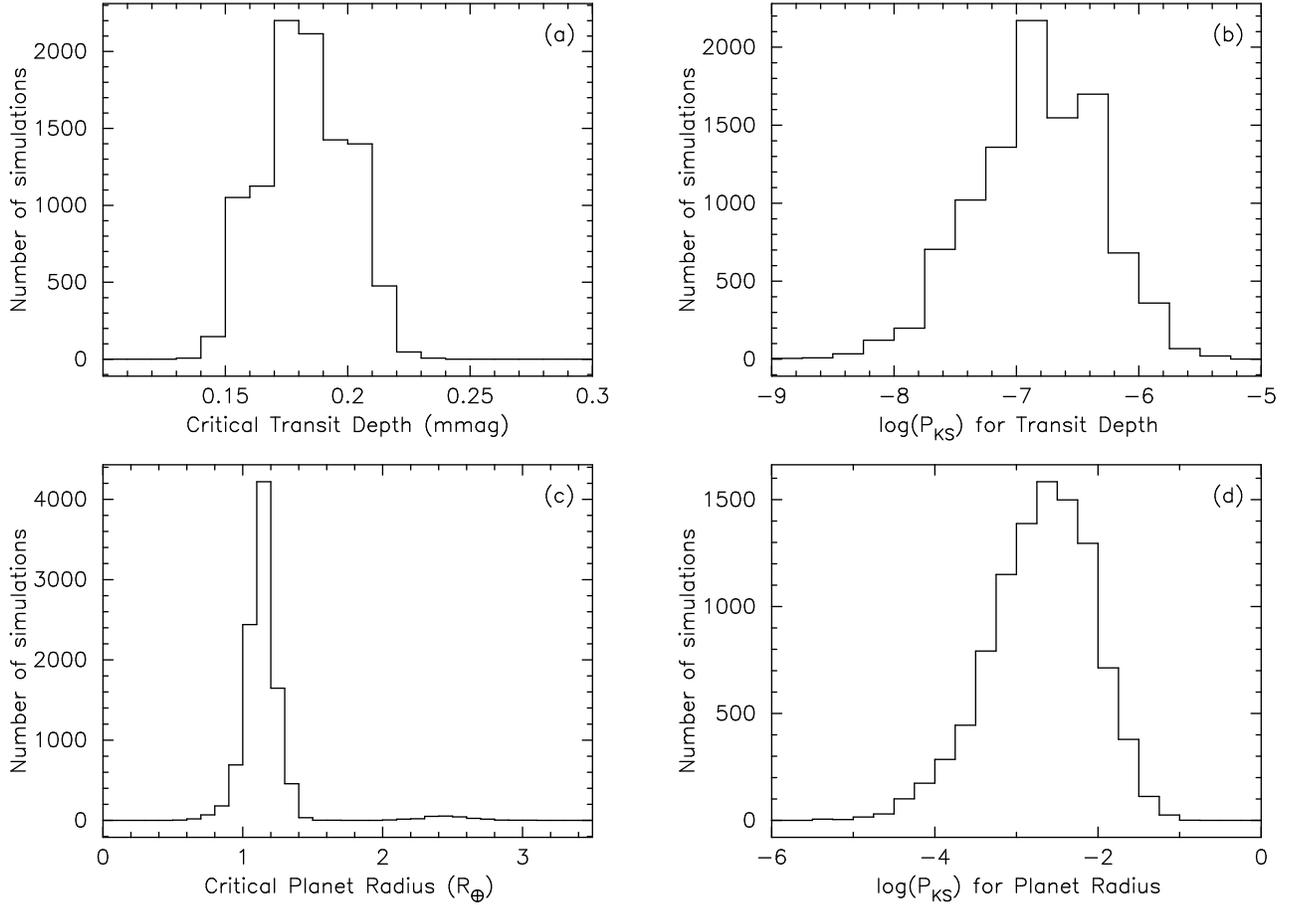}
\caption{
Distributions from a Monte Carlo simulation to test the idea that a
critical transit depth and planet size divides the sample of light
curves by light curve variability level are plotted.  The simulation
investigates the role of uncertainties in the transit depth and planet
radius measurements.  In panel a, a critical transit depth of
$0.18\pm0.02$~mmag divides the sample of host stars into two.  Host
stars showing shallower transits than this have significantly quieter
light curves than the rest of the sample.  Panel b shows the
distribution of the significance of this phenomenon over all K-S
tests.  Similarly, panel c shows a critical planet radius
($1.18\pm0.11$~\re) that divides the sample into two.  Host
stars showing the smallest planets have significantly quieter light
curves than the rest.  Panel d shows the corresponding distribution of
its significance (most simulations show significance at the 99\%
level).}
\end{figure}

\begin{figure}
\includegraphics[angle=-90,scale=0.7,keepaspectratio=true]{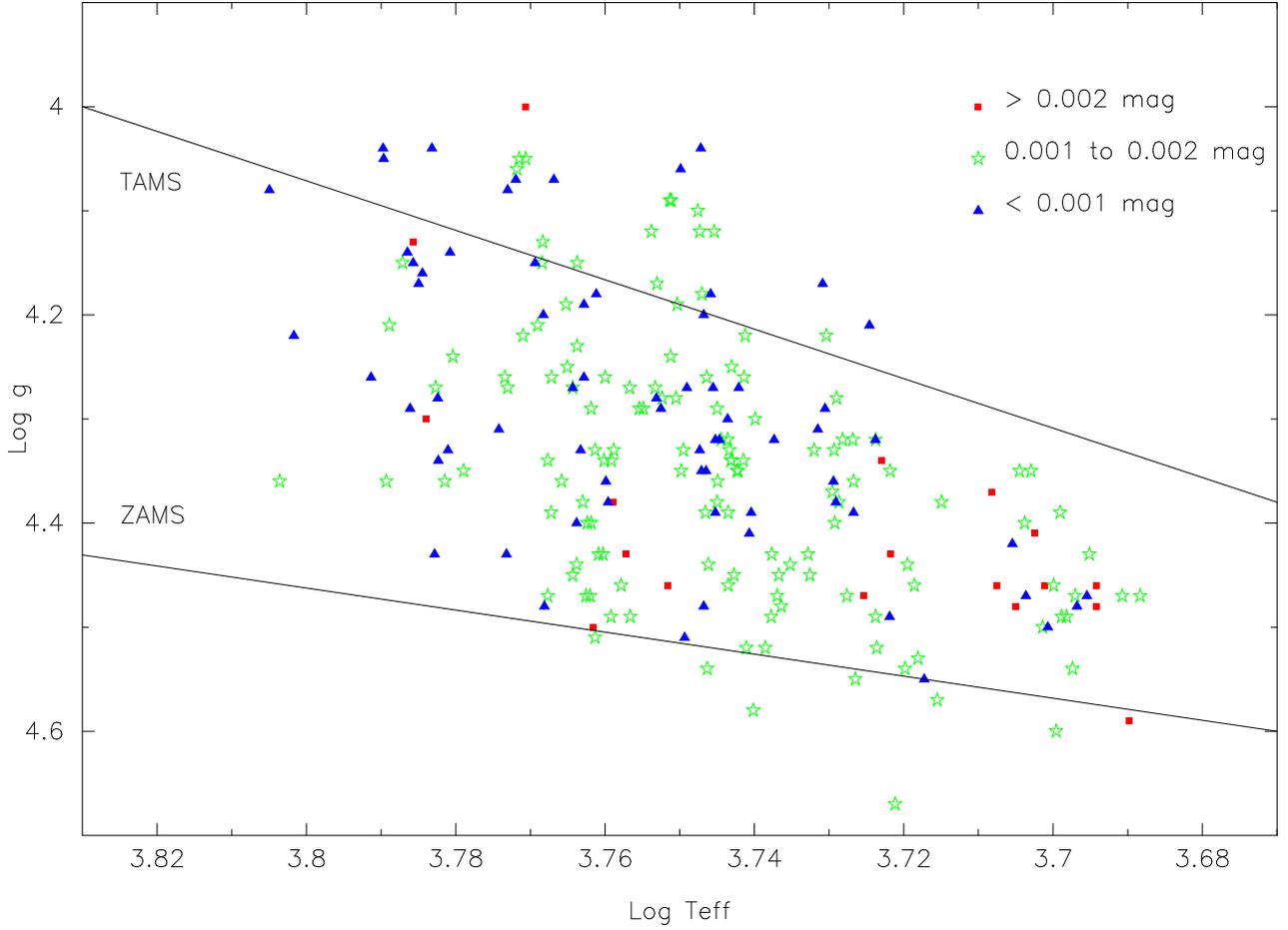}
\caption{Our $Kepler$ small exoplanet host star sample is presented again in this pseudo H-R diagram, 
but now with variability 
information provided for each star. Plotted are 0.25-day light curve standard 
deviations that are $>$0.002 mag (red points), between 0.001 and 0.002 mag
(green points), and those with values $<$0.001 mag (blue points). ZAMS and TAMS lines are shown as described in
Figure 1. 
No apparent relationship between location in the log \teff-log g plane and 
the transit timescale (0.25 day) light curve variability is seen.
}
\end{figure}

\begin{figure}
\plotone{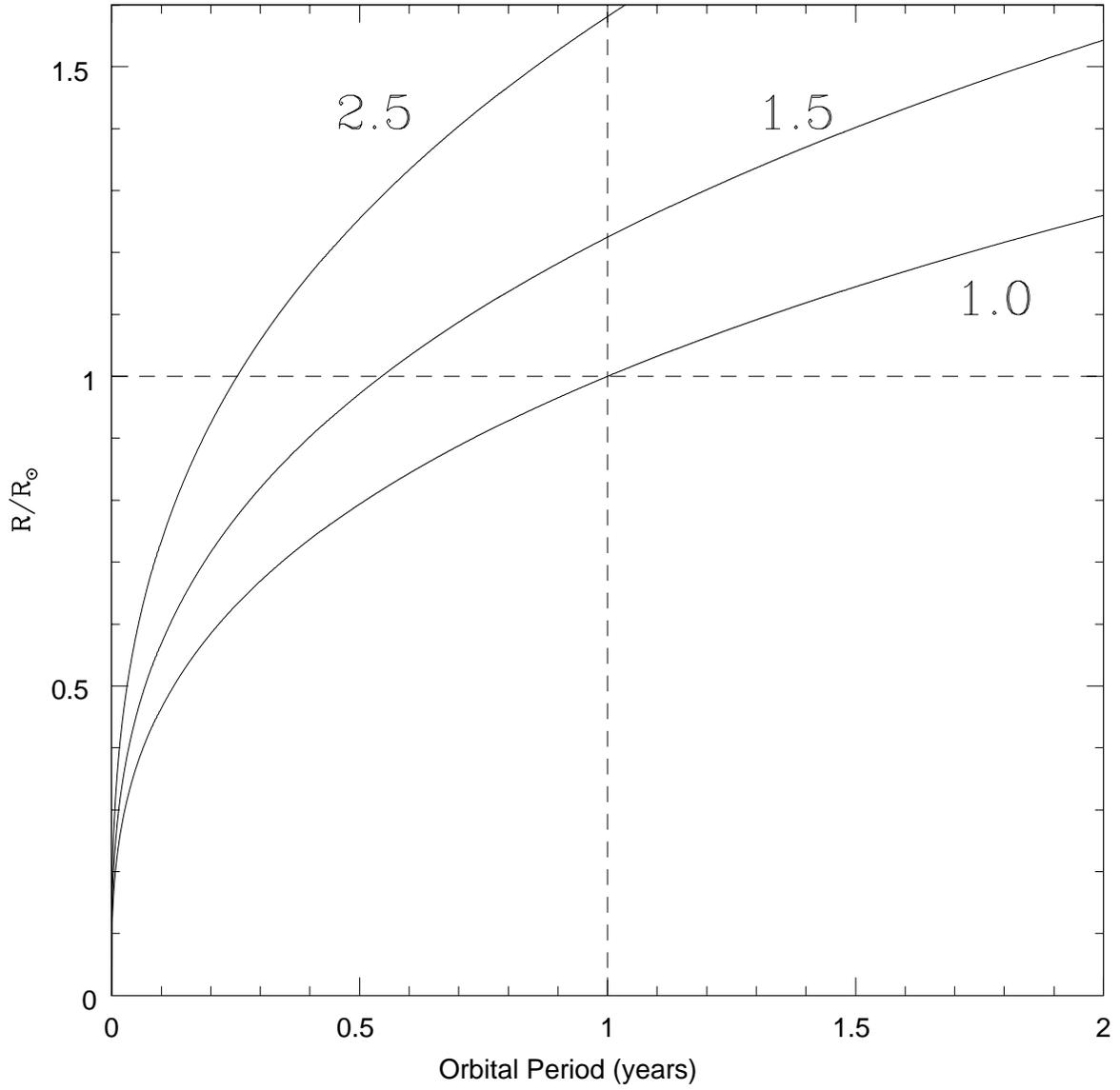}
\caption{Host star radii versus exoplanet orbital period for exoplanet sizes of 1.0,
1.5 and 2.5 earth radii, respectively, as calculated from equation (1).  Dashed lines
at  an orbital period of one year and one solar radius, respectively, are shown for
reference. See \S 3.3 for a discussion.}
\end{figure}


\begin{figure}
\includegraphics[angle=0,scale=0.75,keepaspectratio=true]{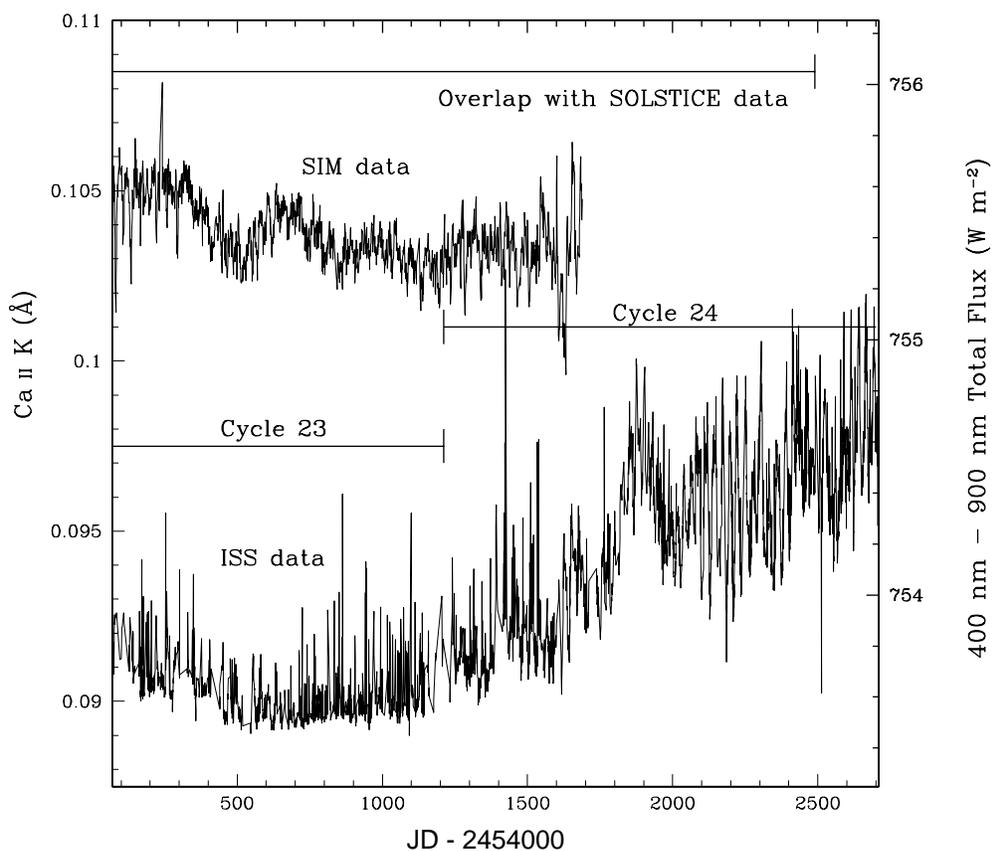}
\caption{The time series of the 1 {\AA} index in the core of the Ca {\sc II} K line as obtained with the SOLIS ISS
on Kitt Peak since 2006 December (JD 2454072).  The K-line parameter is centered at 3933.68 {\AA}.  The ISS data
extend from the declining phase and extended minimum (2008 -- 2010) of Cycle 23 through the rise toward maximum of
the current Cycle 24. The coincident SIM data and the overlap with the time series of data from the SOLSTICE instruments on board the SORCE satellite are indicated.}
\end{figure}

\begin{figure}
\includegraphics[angle=0,scale=0.7,keepaspectratio=true]{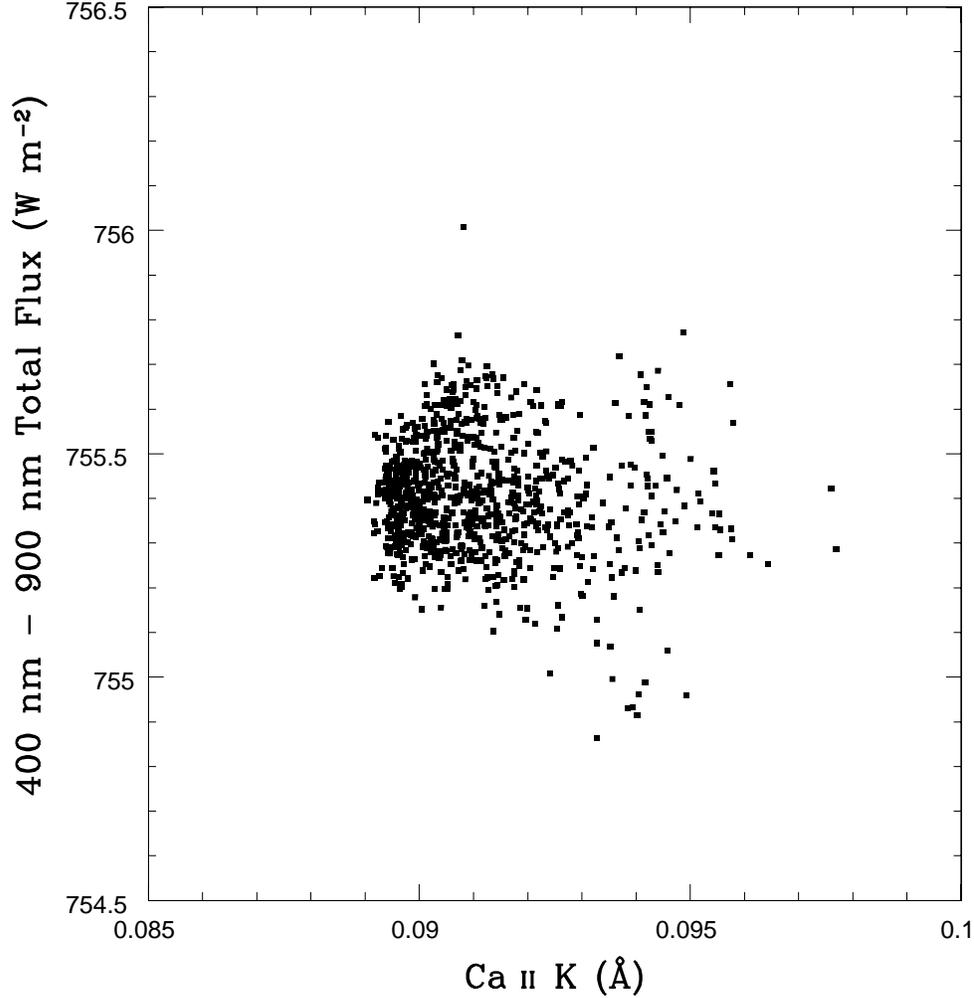}
\caption{The total observed solar flux in the 400 nm - 900 nm visible $Kepler$ bandpass versus the relative
strength of the 1 {\AA} chromospheric Ca {\sc II} K-line parameter centered at 3933.68 {\AA} from
near-simultaneous space- and ground-based data, respectively.  The broadband data were obtained with the SIM
instrument on board the SORCE satellite while the Ca {\sc II} were acquired by the SOLIS ISS instrument on
the same Julian Day number of observation. The range of dates of overlap of these data-sets is approximately
2006 December to 2011 May. Note the absence of any correlation}
\end{figure}

\begin{figure}
\includegraphics[angle=0,scale=0.7,keepaspectratio=true]{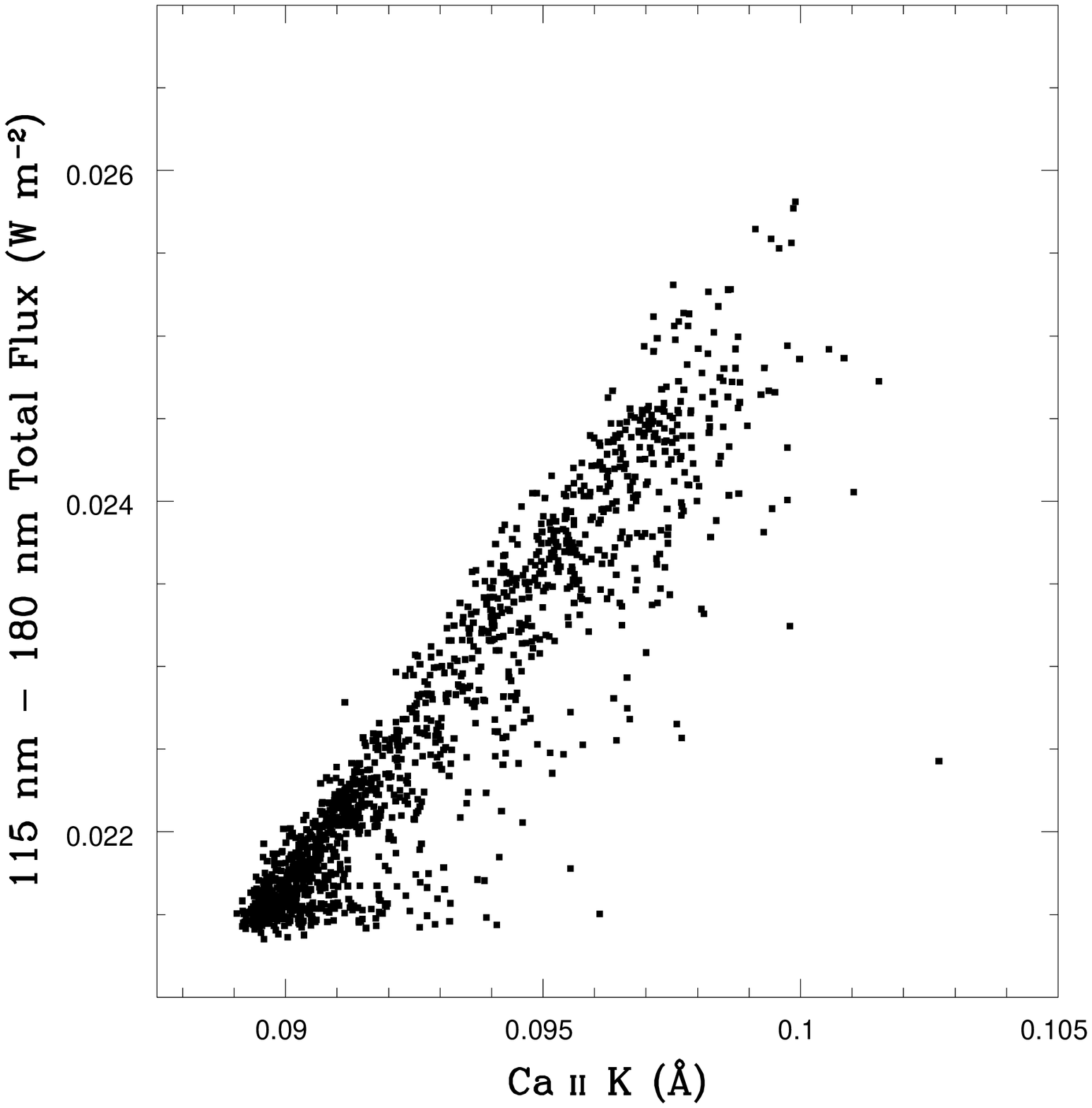}
\caption{The chromospheric Ca II K-index centered at 3933.68 {\AA} and the total observed solar flux in the
far UV bandpass from 115 nm - 180 nm as recorded on the same Julian Day number by the SOLIS ISS instrument
and the SOLSTICE instrument onboard the SORCE satellite. The range of dates of overlap of these data-sets is
approximately 2006 December to 2013 July}
\end{figure} 

\appendix
\section{V Magnitude from J-K Color}

The UBV survey of Everett, Howell, \& Kinemuchi (2012) covered most of the $Kepler$ field and provided U, B, and V
Johnson magnitudes for over 4 million objects in the field.  The astrometry of the UBV survey is typically
reliable to better than 0.1$^{\prime\prime}$, enabling matching of the KOIs to the UBV catalog - this spatial
matching was performed for all KOIs on the Community Follow-Up Observation Program Website (CFOP;
https://cfop.ipac.caltech.edu) which is part of the NASA Exoplanet Archive at the NASA Exoplanet Science Institute.
However, approximately 5\% of the 6100+ KOIs either did not have a spatial match (within 1$^{\prime\prime}$ to the
UBV catalog or the V magnitude of the UBV magnitude was not consistent with the listed $Kepler$ magnitude or the
other measured magnitudes from the $Kepler$ Input Catalog $(K_P,g,r,i,z,J,H,Ks)$.

For these sources, we have derived a $V-Ks$ vs. $J-Ks$ empirical relationship in order to obtain a V magnitude
estimate for those KOIs that have either no UBV match or the match is suspect $(|K_P - Vmag|>1)$.  In Figure
10  below, $V-Ks$ vs $J-Ks$ is plotted for the 5626 KOIs that have measured $J$, $Ks$, and $V$ magnitudes, and the
relationship was fitted with a 3rd-order polynomial of the form

\begin{equation}
V-Ks = 7.11273(J-Ks)^3 - 8.51190(J-Ks)^2 + 6.33950(J-Ks) + 0.200901
\end{equation}

for -0.5 $\le$ $J-Ks$ $\le$ 1.0 mag.

Comparing the polynomial-derived V magnitudes to the measured V magnitudes (see bottom Figure 8), the magnitudes
agree with a median difference of 

\begin{equation}
\langle V_{poly} - V_{true} \rangle  = 0.002 \pm 0.155\ \rm{mag.}
\end{equation}

The V magnitudes 
on the CFOP website, and those used here, are either the direct spatial match with angular
separation of the UBV source and the KOI listed or the above polynomial estimation. 

\begin{figure}
\includegraphics[angle=0,scale=0.7,keepaspectratio=true]{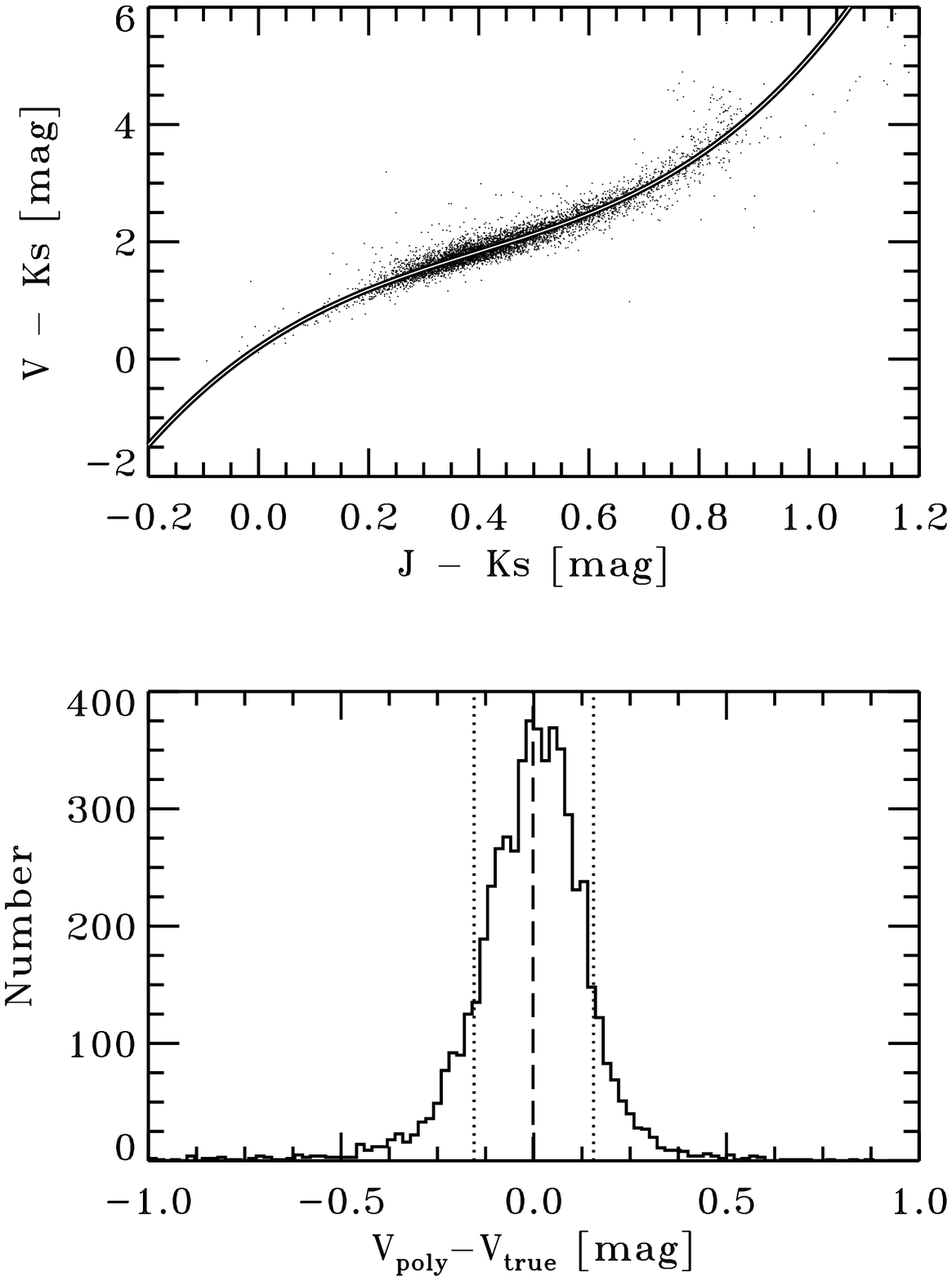}
\caption{
{\it Top:} Plot of $V-Ks$ vs. $J-Ks$ for all 5626 KOIs that have a measured 
V, J, and Ks magnitude.  The over-plotted line represents the fitted 3rd order polynomial.  {\it Bottom:} The
histogram of the residuals from the polynomial fit is shown with the median difference of 0.002 mag marked by the
vertical dashed line and the $1\sigma$ dispersion of 0.155 mag marked by the vertical dotted lines.
}
\end{figure} 

\end{document}